# ESCM: An Efficient and Secure Communication Mechanism for UAV Networks

Haoxiang Luo, *Graduate Student Member*, *IEEE*, Yifan Wu, Gang Sun, *Senior Member*, *IEEE*,
Hongfang Yu, *Member*, *IEEE*, Mohsen Guizani, *Life Fellow*, *IEEE*

*Abstract*—UAV (unmanned aerial vehicle) is rapidly gaining traction in various human activities and has become an integral component of the satellite-air-ground-sea (SAGS) integrated network. As high-speed moving objects, UAVs not only have extremely strict requirements for communication delay, but also cannot be maliciously controlled as a weapon by the attacker. Therefore, an efficient and secure communication method designed for UAV networks is necessary. We propose a communication mechanism ESCM. For high efficiency, ESCM provides a routing protocol based on the artificial bee colony (ABC) algorithm to accelerate communications between UAVs. Meanwhile, we use blockchain to guarantee the security of UAV networks. However, blockchain has unstable links in high-mobility networks resulting in low consensus efficiency and high communication overhead. Consequently, ESCM introduces digital twin (DT), which transforms the UAV network into a static network by mapping UAVs from the physical world into Cyberspace. This virtual UAV network is called CyberUAV. Then, in CyberUAV, we design a blockchain consensus based on network coding, named Proof of Network Coding (PoNC). Analysis and simulation show that the above modules in ESCM have advantages over existing schemes. Through ablation studies, we demonstrate that these modules are indispensable for efficient and secure communication of UAV networks.

*Index Terms*—UAV networks, routing protocol, digital twin, blockchain consensus, network coding

## I. INTRODUCTION

IN recent years, the rapid development of communication technology has helped UAVs show great potential in the public and civil fields, and it has also become the focus of attention in the communication field [1]. For example, Ranjha *et al.* [2] consider using UAVs as decode-and-forward relays to communicate between controllers and multi-mobile robots. In [3], the authors study the quality-of-service capability of UAVs in mobile edge networks. Moreover, as an integral part of the SAGS network, UAVs can assist 6G communications by performing indispensable functions such as auxiliary computing and communication relaying. In [4], the authors use UAVs-assisted mobile edge computing systems to achieve the above functions. In [5], the authors design a multi-user communication network coordinated by UAVs and base stations.

However, there are still many problems that need to be solved before the UAV network can achieve efficient and reliable functions. Although there is much work in the area of mobile ad hoc networks, and vehicular ad hoc networks, it does not address the unique characteristics of the UAV network. The faster dynamic changes than vehicles and the fluid topologies both make the UAV network topology change more frequently and unpredictably. As a result, we need to design specific communication mechanisms for the UAV network.

*A. Research Motivation*

To adapt to the fast-moving characteristics of drones and meet the high efficiency of UAV network communication, many scholars design reasonable and efficient routing protocols for UAV networks to improve communication performance and efficiency. For example, the routing protocols are based on geographical location proposed by [6-7]. In addition, there are routing protocols based on UAV network topology, such as [8-9]. Recently, bio-inspired methods have received great attention due to their potential properties, such as adaptability, self-organization, and robustness [10]. Since the UAV fleet and swarm are similar in that they both fly at high speed in three-dimensional space, thus, ABC [11] is selected as the improved method of routing protocol in ESCM. This routing protocol will select several suitable drones as candidate relay nodes, which will play a role and determine the most suitable relay node in the PoNC consensus proposed by us.

However, the routing protocol can only improve the communication efficiency of the UAV network, but cannot solve the existing communication security problems in the UAV network as described in [12]. Currently, emerging blockchain technology provides a viable solution to security issues of the UAV network. It can effectively avoid many security problems of traditional encryption algorithms, such as single-point failure [13-14]. Blockchain is already playing a huge role in the Internet of Energy (IoE) [15] and the Internet of Vehicles (IoV) [16] through consensus algorithms that audit and record all nodes inforamtion in the network. There is also a lot of research on blockchain in UAV networks. For example,

This work was supported by the Natural Science Foundation of Sichuan Province under Grant 2022NSFSC0913. (Corresponding author: *Gang Sun*, *Hongfang Yu*.)

Haoxiang Luo, and Gang Sun are with the Key Laboratory of Optical Fiber Sensing and Communications (Ministry of Education), University of Electronic Science and Technology of China, Chengdu 611731, China (e-mail: lhx991115@163.com; gangsun@uestc.edu.cn).

Yifan Wu is with the Department of Electrical and Computer Engineering, Carnegie Mellon University, Pittsburgh, PA 15260, USA (e-mail: yifanwu2@andrew.cmu.edu).

Hongfang Yu is with the Key Laboratory of Optical Fiber Sensing and Communications (Ministry of Education), University of Electronic Science and Technology of China, Chengdu 611731, China, and also with Peng Cheng Laboratory, Shenzhen 518066, China (e-mail: yuhf@uestc.edu.cn).

Mohsen Guizani is with with the Machine Learning Department, Mohamed Bin Zayed University of Artificial Intelligence (MBZUAI), Abu Dhabi, United Arab Emirates (e-mail: mguizani@ieee.org).

in [17], the authors propose a blockchain-based trusted framework for UAV networks, the corresponding network architecture, protocol stack, and key control method. In [18], the authors present a case of the UAV network based on blockchain and federated learning in 6G.

As an important part of a blockchain, the consensus determines the operational efficiency of the blockchain system to some extent [19]. Therefore, some scholars also study consensus algorithms used in the blockchain for UAV networks [20]. However, the implementation of consensus is related to participating nodes. In the UAV network, new drones often join the formation or some drones leave the formation, resulting in the instability of blockchain connection and seriously affecting the consensus efficiency. Moreover, reaching consensus requires many drones to carry out multiple rounds of physical-to-physical communications (P2PC), thus dragging down the efficiency of communication [21-22].

The rapid development of DT technology offers a solution to the above problems. By building digital replicas, physical entities can be mapped into Cyberspace. In this case, traditional P2PC can be converted to virtual-to-virtual communication (V2VC) via edge network and core network [23]. More importantly, replicas in Cyberspace constitute a static network that is not limited by the geographic location of UAVs, showing great potential in the management of drones. In order to explore the performance of the UAV network in Cyberspace, we introduce this concept in ESCM, and map the highly mobile UAV network into a static network in Cyberspace, namely CyberUAV.

Then, we consider the throughput is also an important indicator of communication performance in the mapped CyberUAV, thus, we design a consensus algorithm PoNC based on network coding to increase the throughput of UAV networks. PoNC is a consensus algorithm similar to PoW [24], but it replaces meaningless mining behavior in PoW with meaningful coding behavior. PoNC chooses the most suitable drone as a relay node to encode and forward the transmitted message by comparing the coding capability of candidate drones selected in the ABC routing protocol.

As mentioned above, communication difficulties in current UAV networks can be summarized into the following three points, which are also problems that ESCM focuses on solving:

- How to design efficient routing protocol and consensus algorithm to ensure the timeliness of UAV network communication.
- How to use blockchain technology to ensure the security of the UAV network.
- How to solve the unstable link problem of blockchain in the high-speed mobile environment of UAV network.

### B. Our Contributions

- To promote the message transmission rate and consensus efficiency, this work adopts ABC to improve the routing protocol, by selecting several suitable candidate drones as candidate relay nodes to define the scope for the subsequent blockchain consensus, which can avoid global negotiation in the UAV network and unnecessary communication costs.
- Additionally, to avoid the instability of blockchain links due to the high-speed movement of drones, we introduce the concept of the DT in ESCM, mapping drones in the physical world to Cyberspace, and transforming the high-speed moving UAV network into a static network named CyberUAV. Then, we introduce blockchain into static CyberUAV to improve the security of the UAV network.
- To further improve consensus efficiency and increase throughput, based on the blockchain system consisting of static networks in CyberUAV, we propose a consensus algorithm based on network coding, named PoNC, to further improve network throughput. It can replace meaningless mining behavior in PoW with meaningful coding behavior. Simulation results show that it not only increases throughput and improves consensus efficiency, but also ensures high security.
- We theoretically derive and simulate the consensus success rate of MobileChain (blockchain in a dynamic wireless network). It is proved that the static CyberUAV in DT has a higher consensus success rate than that of mobile UAV networks. This result further strengthens our motivation to use DT to enable MobileChain.

### C. Structure of this Paper

The remaining contents are arranged as follows. Section II reviews the related work. Section III shows the derivation and simulation of the MobileChain consensus success rate, leading us to enable MobileChain with DT. Section IV introduces the ESCM model. Section V describes details of the ABC routing, construction of CyberUAV, coding scheme, and PoNC consensus. Section VI presents the performance analysis, including the consensus overhead and consensus security of PoNC. Section VII is the performance evaluation of ESCM. Section VIII gives the conclusion of this work.

## II. RELATED WORK

### A. Routing Protocol in UAV Networks

To improve the communication efficiency of UAV networks, many researchers choose to design routing protocols. On the one hand, many scholars have proposed routing protocols based on geographical locations. In [6], the authors propose the geographic position based on hopeless opportunistic routing. It combines the advantages of hopeless protocol and location-based protocol to make full use of the transmission opportunities of all links. Vikramjit *et al.* [7] design an adaptive beaconing scheme to improve the accuracy of the neighbor table for geographical routing in UAV networks. On the other hand, there is another routing protocol based on network topology. For example, Li *et al.* [8] establish a reference model and a statistical model of the communication channel in the UAV network topology.

The routing protocols listed above belong to the classical computing domain, where mathematical ideas are considered to obtain provably optimal routing solutions. However, such solutions are often not suitable for large-scale UAV networks because of the characteristics of strong time-varying network topology and high-speed movement, which will reduce the success rate of packets transmitted. Recently, a method called bionics has received a lot of attention because of its applicability to a variety of applications. It is produced by imitating the laws of natural species, with excellent performance and outstanding ability to solve theoretical and experimental problems, such as ant colony algorithm [25]. A typical one is that Zhang *et al.* [26] propose a 3D transformative routing protocol according to the human nervous system, including proactive routing skeleton establishment, reactive path selection, and bottleneck-aware route maintenance.

Based on the advantages of the above bionic algorithm, we plan to take ABC as the routing method of ESCM. However, the design of routing alone cannot avoid communication security issues of UAV networks. Therefore, we consider using ABC to select some UAVs with great communication performance as candidate relay nodes for transmitting information, and select the best relay node in these nodes through blockchain consensus to improve security.

*B. Blockchain Technology in UAV Networks*

As a comprehensive solution to security problems, blockchain has also been taken by many scholars as an important method to improve the security of UAV network communication. Mehta *et al.* [27] review the roles and challenges of blockchain in UAV networks, showing that blockchain can make communication between UAVs secure. In [28], blockchain is deployed within drones, supported by fog computing and the data center to provide a secure authentication service and data availability.

In addition, due to the important role of consensus algorithm in a blockchain system, many scholars also pay attention to blockchain consensus design in UAV networks. The authors from [12] design a consensus algorithm for UAV networks from the perspective of game theory, which has a low communication overhead. Wang *et al.* [19] propose an improved credibility enhancement PBFT consensus and a credibility assessment scheme based on the reliable records of UAV behaviors. According to the characteristics of UAV networks, in [20], the authors propose a UAV consensus Protocol (UCP), which includes a message exchange stage and a consensus stage.

The above-mentioned literature is only on how to cooperate with the high-speed mobility of drones and the uncertainty of the communication links to design consensus algorithms for UAV networks. However, our ESCM considers that the UAV network is converted into a static network to design a simple and effective consensus algorithm. Then, it also addresses the security and connectivity instability of the UAV networks. In addition, no scholars have considered designing blockchain consensus from the network coding perspective to increase throughput in UAV networks. Therefore, we consider designing a consensus algorithm PoNC based on network coding according to the static network in Cyberspace.

*C. Digital Twin for UAV Networks*

The DT model is a digital representation of drones in the application layer, which can model and service the whole life cycle of drones [29]. Using this technology to solve the communication problem of high-speed moving objects has gradually become an academic hotspot, such as [22]. Meanwhile, in [30], after mapping a UAV network to Cyberspace, authors combine federated learning to solve the contradiction between privacy protection and data training in the network. In [31], the authors design a method combining DT with convolutional neural networks in UAV networks, which has great security performance. Shen *et al.* [32] propose a training framework for deep reinforcement learning (DRL) based on DT. This DRL model can be learned from Cyberspace and quickly deployed to the real-world UAV network with the help of Cyberspace.

To summarize, existing literature mainly focuses on improving some performance of UAV networks in Cyberspace using different artificial intelligence algorithms. Only Dai *et al.* [33] have proposed a secure communication scheme for blockchain-enabled air-ground integrated networks with DT. Consequently, the combination of a UAV network with blockchain technology in Cyberspace still needs further investigation. In particular, the DT-enabled mobile blockchain can significantly improve wireless communication stability in blockchain consensus.

III. CONSENSUS SUCCESS RATE OF MOBILECHAIN

This section presents the derivation and simulation of the MobileChain consensus success rate, demonstrating the motivation we need to use DT to enable MobileChain.

Before studying the consensus success rate, we need to be clear that most consensus depends on multiple-round communications, such as Practical Byzantine Fault Tolerance (PBFT) [19], and our PoNC consensus (more details in Section V-D). Therefore, we can take the success rate of each communication as an important indicator of the consensus success rate.

We assume that the initial UAV network is on the ground, and $k$ candidate relay drones obey a two-dimensional Poisson distribution with density $\gamma$ in an area with radius $R$. According to the two-dimensional Poisson distribution, the probability density function of the distance $r$ between two drones is

$$f(r) = \frac{d(r^2/R^2)}{dr} = \frac{2r}{R^2}. \quad (1)$$

When the channels in this UAV network conform to Rayleigh fading (RF), the Signal to Noise Ratio (SNR) is

$$SNR = \frac{P_T h r^{-\alpha}}{P_N}, \quad (2)$$

where $P_T$ is the node's transmit power; $h$ represents a non-negative random variable of power gain in RF, which follows a negative exponential distribution with exponent 1. $\alpha$ represents the path loss exponent; $P_N$ is the interference noise power.

Then, we set the SNR threshold at which drones can recover signals as $z$, then according to the two-dimensional Poisson distribution [34], the average success rate of communications in a static wireless environment is

$$P_s = \int_0^R P\{SNR > z\} f(r) dr = \frac{2\pi\gamma}{k} \int_0^{\sqrt{k/(\pi\gamma)}} \exp\left\{\frac{-P_N r^\alpha z}{P_T}\right\} r dr. \quad (3)$$

However, the UAV network is a dynamic network, meaning that the distance $r$ can change at any time, and the Doppler effect from motion can also affect the power $P_T$. Therefore, in order to explore the transmission success rate in MobileChain, we need to analyze $r$ and $P_T$ in a dynamic situation separately.

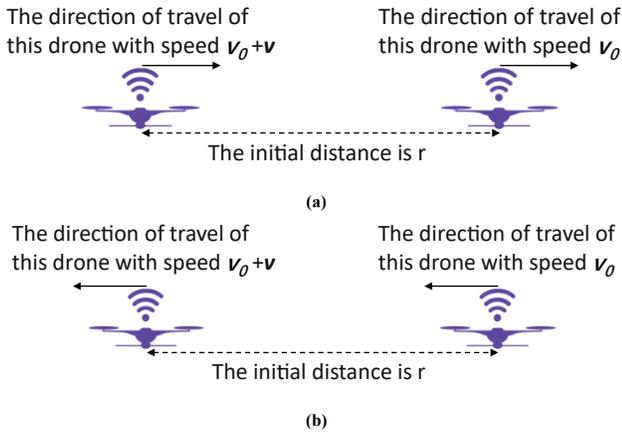

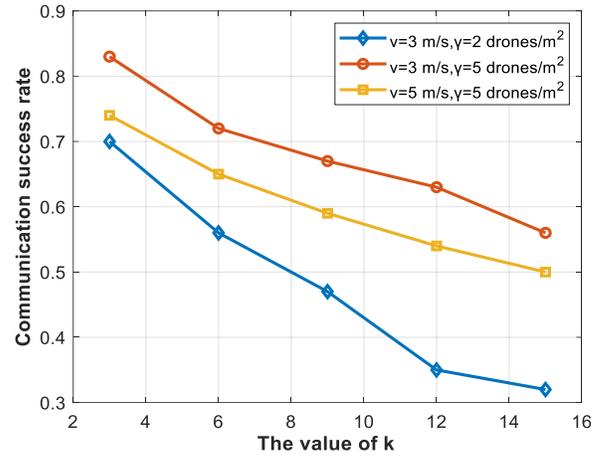

**Fig. 1.** (a) Relative distance variation with -*vt*; (b) Relative distance variation with *vt*.

**Fig. 2.** Communication success rate in MobileChain.

First, we analyze the change in distance. If the relative velocity between drones is $v$, then the relative distance between the two drones will change $vt$ after $t$ seconds. Further, due to the directionality of motion, the lower bound of relative distance change is $-vt$ (as shown in Fig. 1a). The upper bound of the relative distance change is $vt$ (as shown in Fig. 1b). As a result, the relative distance between drones is

$$r_m = |r + d|,$$
$$d \in [-vt, vt]. \qquad (4)$$

Second, we analyze the change in power. The power change caused by the movement is mainly due to the Doppler effect generated, which will further attenuate the signal power in a specific channel environment. According to [35], in our described Rayleigh channel, the Doppler power spectral density function of the signal is

$$S(f) = \begin{cases} \dfrac{P_T^2}{\pi f_m \sqrt{1-(f/f_m)^2}}, & |f| \le f_m \\ 0, & |f| > f_m \end{cases} \qquad (5)$$

where $f_m$ is the maximum Doppler-frequency shift, and $f$ is the original frequency of the signal. According to the definition of the maximum Doppler shift, we have

$$f_m = \dfrac{v}{c} f, \qquad (6)$$

where $c$ represents the light speed.

Furthermore, based on the Wiener-Khinchin theorem [36], we find the relation between the autocorrelation function and power spectral density is

$$R(\tau) = \int S(f) e^{j2\pi f\tau} df. \qquad (7)$$

If the signal transmitted in the UAV network satisfies the wide-sense stationary process, when $\tau=0$, $R(0)$ equals the average power $P_{av}$ of the signal, namely

$$R(0) = \int S(f) df = P_{av}. \qquad (8)$$

Thus, under the influence of the Doppler effect in the Rayleigh channel, the average power of the signal can be expressed as

$$P_{av} = \dfrac{P_T^2}{\pi} \sin^{-1}\left(\dfrac{c}{v}\right). \qquad (9)$$

As a consequence, in the mobile UAV network, the communication success rate is

$$P_{s\_m} = \dfrac{2\pi\gamma}{k} \int_0^{\sqrt{k/(\pi\gamma)}} \exp\left\{\dfrac{-P_N |r+d|^\alpha z}{P_{av}}\right\} r\, dr. \qquad (10)$$

Now, we assume that $P_T = 0.5$ W, $P_N = 0.1$ W, $\alpha = 2$, $t = 10$ s, $z = 6$dB, and set initial three groups of $v$ and $\gamma$, including $v=3$ m/s, $\gamma=2$ drones/m²; $v=3$ m/s, $\gamma=5$ drones/m²; $v=5$ m/s, $\gamma=5$ drones/m². Then, we simulate the consensus success rate according to these values.

The results are shown in Fig. 2, and we can find the increase of $k$, $v$, and the decrease of $\gamma$ have a negative impact on the communication success rate. Additionally, these simulation results show that in the wireless mobile scenario, communications are greatly affected. However, in the virtual space constructed by DT, communications are not affected by various channel fading and path loss from the physical world, and the communication success rate is almost 100% [21]. Therefore, this rule proves the superiority of the PoNC consensus process in CyberUAV, to avoid a low consensus success rate in MobileChain.

IV. METHOD OVERVIEW OF ESCM

This section describes the model framework of ESCM, including ABC routing, CyberUAV virtual network, and PoNC consensus. Fig. 3 shows the overall process. Moreover, we also introduce the adversary model of ESCM.

*A. Model Framework*

First, it is assumed that *drone A* needs to send *message A* to *drone B*. However, due to the limitation of distance and speed, *drone A* cannot randomly select other drones as relay nodes to relay the message. Therefore, *message A* needs a proper communication path, namely, a reasonable routing protocol. It should be noted that although the ABC algorithm is used to improve the routing protocol, the unique feature of ESCM is that it does not directly determine the final communication path according to the routing protocol. Instead, the ABC algorithm finds several suitable candidate relay nodes and turns them over to the subsequent PoNC consensus to determine the best relay node. This design can avoid the security problem in that the routing protocol selects a single path to transmit messages, which leads to interception by the attacker.

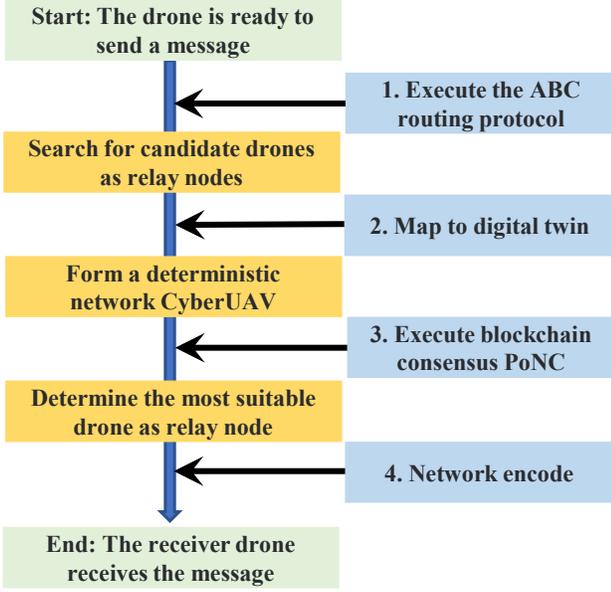

**Fig. 3.** The process of ESCM.

Second, ESCM maps the sender *drone A*, the receiver *drone B*, and several candidate relay nodes selected by the routing protocol to Cyberspace with the help of the concept of DT. And then, the virtual network of drones in Cyberspace is called CyberUAV. Since nodes in the virtual world are only digital replicas of drones in the physical world and do not fly as fast as they do, CyberUAV is a static network.

Third, after the formation of CyberUAV, Cyberspace runs a blockchain process with the proposed PoNC consensus to determine the best relay node according to the coding ability of candidate relay nodes. The PoNC consensus proposed by us imitates the design of PoW consensus, but PoNC replaces meaningless mining behavior in PoW with meaningful coding behavior, avoiding wasting computing resources. Through such a consensus process, ESCM can not only determine the relay node with the best encoding effect, but also be recognized by other nodes in the consensus, ensuring the authority and security of the process.

Finally, CyberUAV informs the consensus result to the physical world, and the message is forwarded to the receiver drone by the optimal relay node determined by PoNC.

### B. Adversary Model

Although ESCM is enabled by blockchain, it may also be attacked by two attacks. We divide attacks into two types, one external and the other internal.

*1) External Attack:* For the external attack, we establish a model for a double spend attack (DSA), that is, the attacker replaces the honest nodes (HNs) in PoNC by generating blocks through malicious nodes (MNs), and finally controls the blockchain.

*2) Internal Attack:* For the internal attack, we consider that when mapping consensus nodes, malicious drones (MDs) are mapped to CyberUAV to form MNs, which misleads the consensus result, causing the consensus to fail.

The MNs of the internal attack are different from those of the external attack, where the MNs of the external attack are accessed by the attacker, while the MNs of the internal attack indicates that consensus nodes are malicious.

## V. DESIGN DETAILS OF ESCM

In this section, ABC routing protocol, construction of CyberUAV, network coding, and PoNC consensus are described in detail, respectively.

### A. ABC Routing Protocol

Before explaining the routing protocol based on ABC, it is necessary to introduce the ABC algorithm.

*1) ABC Algorithm*

According to [37], the ABC algorithm divides bees into three roles: employed bee (EB), onlooker, and scout. EBs and onlookers are responsible for mining food, and scouts are responsible for exploring food. The specific process of the algorithm is as follows:

In the initialization stage, a uniformly distributed initial population containing $SN$ food sources is randomly generated, the dimension of each food source $x_i$ ($i=1, 2, ..., SN$) is $D$, and each solution is initialized by (11).

$$x_{ij} = x_{\min,j} + rand(x_{\max,j} - x_{\min,j}), \quad (11)$$

where, $x_i$ is the *ith* solution of the population, $x_{min,j}$ and $x_{max,j}$ are the lower bound and upper bound of the *jth* dimension respectively, $j \in [1, 2, ..., D]$, and $rand$ is the random number in $[0,1]$.

In the EBs stage, populations explored new food sources according to (12).

$$v_{ij} = x_{ij} + \phi_{ij}(x_{ij} - x_{kj}), \quad (12)$$

where, $k \in [1, 2, ..., SN]$, $k \neq i$ and the $\phi_{ij}$ is a random number in $[-1, 1]$. After the new solution is produced, a better food source is retained between $x_i$ and the new solution $v_i$. The fitness for each food source $fit_i$ is calculated as follows:

$$fit_i = \begin{cases} \dfrac{1}{1+F(x_i)}, & F(x_i) \geq 0 \\ 1+|F(x_i)|, & F(x_i) < 0 \end{cases}, \quad (13)$$

where, $F(x_i)$ represents the value of the objective function about $x_i$. After mining the food source, employed bees return to the hive and share the location and information of the food source with onlookers through a special dance. In the onlooker's stage, onlookers calculate the selection probability $p_i$ of food source $x_i$ based on the information shared by employed Bees.

$$p_i = 0.9 \cdot \dfrac{fit_i}{\max(fit_i)} + 0.1. \quad (14)$$

Onlookers select food sources by roulette and updated according to (12). If any food source has been mined for *limit* times, then the food source is abandoned, and the corresponding EM will transform into the scout to explore a new food source. The process of the ABC algorithm is shown in Algorithm 1.

**Algorithm 1: ABC Algorithm**

| | |
|---|---|
| 1 | **Initialization:** *SN*, the maximum number of iterations *maxGen*, current number of iterations *Gen*=1, and *limit* |
| 2 | **Choose the food source (BM's stage):** |
| 3 | **Input:** $x_i$, $k$, $\phi_i$ |
| 4 | **Search for food sources:** Calculate the value of $v_i$ according to (12) |
| 5 | **Calculate fitness values for new food sources:** Calculate the value of $fit_i$ according to (13) |
| 6 | **Choose a food source with good fitness values:** By comparing the fitness of $x_i$ and $v_i$ |
| 7 | **Output:** The food source |
| 8 | **Update food sources (onlookers):** |
| 9 | **Input:** Fitness of each food source |
| 10 | **Calculate the selection probability of onlookers:** According to (14) |
| 11 | **Create new food sources:** According to (12) |
| 12 | **Determine whether to keep the food source:** By comparing the fitness of old and new food sources |
| 13 | **Output: The better food source** |
| 14 | **Search for new food sources (Scout):** |
| 15 | **if** mining times of food source = *limit*, **do** |
| 16 | The corresponding EM transforms into the scout to explore a new food source |
| 17 | **else** continue to mining |
| 18 | **end if** |
| 19 | **if** *Gen=maxGen*, do |
| 20 | Output the optimal solution |
| 21 | **else** return to the second line |
| 22 | **end if** |

*2) ABC Routing Protocol*

In this protocol, the UAV network is compared to a bee population, and drones are regarded as the food source. Drones in the network will periodically generate messages containing the communication range, which are regarded as bee messages (BM). Each step of the bee's progress is considered a hop transfer of BM from the current vehicle to the next. Each drone has two state parameters: equipment loads state $\theta$ and transmission success rate $\mu$. The food concentration can be calculated by (15), where *a* and *b* are weights of equipment load and transmission success rate respectively, satisfying *a*+*b*=1.

$$F(x_i) = a\theta + b\mu. \qquad (15)$$

In the beginning, all employed BMs (EBMs) are scout BMs (SBMs) without prior knowledge. When an SBM stays in drone $x_1$ and finds other suitable food drones according to (12), it calculates the drone's fitness. If the fitness of $x_2$ is the optimal solution according to (13) and (15), it will be the best food drone for transmission. And then, the SBM turns into an EBM and begins to eat (travel to $x_2$).

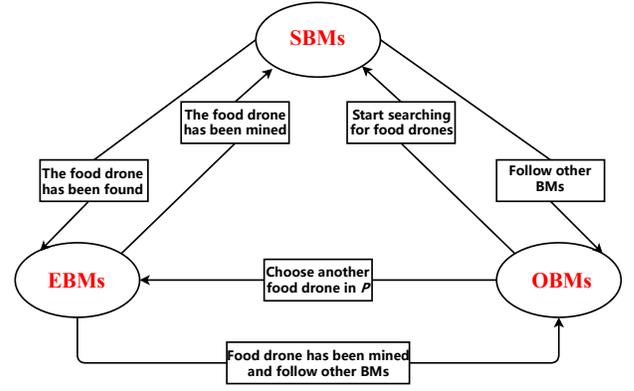

**Fig. 4.** The transformation relationship among EBM, SBM and OBM.

We consider the reality of routing protocols in the UAV network, which is unlikely that EBMs return to the hive and notify onlook BMs (OBMs) the best food drone. It will greatly increase communication overhead and latency. Therefore, when an EBM finds $x_2$, we consider storing $x_2$ in the priority set *P* which is a set that represents the next hop of $x_1$, to inform subsequent OBMs conveniently. Then, we set the crowding factor $\eta$ to be placed in $x_2$. The crowding factor is the number of BM arriving at the food drone. If $\eta<limit$, the EBM stays in this food drone; If $\eta=limit$, it indicates that the food drone has been mined, and the EBM converts to an SBM and continues the above steps.

Furthermore, there is a step for the OBM. When an OBM reaches $x_1$, calculates the selection probability of food drones stored in set *P* according to (14). And then the OBM selects the food drone by roulette. If it is $x_2$, the OBM should update the $\eta$ value of this food drone. Additionally, it is necessary to determine whether to replace the food drone by judging the values of the $\eta$ and *limit*. If $\eta<limit$, the OBM remains in this food drone; If $\eta=limit$, the OBM will delete $x_2$ from set *P*, and choose the other food drones stored in set *P*. Maybe it is $x_3$, which satisfies $\eta<limit$. Moreover, the OBM will transform into an EBM when it selects other food drones in *P*. If it doesn't find a better food drone in *P*, it will convert to an SBM.

Therefore, through the above steps, BMs can find candidate UAVs as relay nodes for the next hop transmission and place them into the set *P*. Among them, the transformation relationship among EBMs, SBMs, and OBMs is shown in Fig. 4.

And the ABC routing protocol brings the following two benefits to ESCM:
- First, as an efficient bionic algorithm, ABC can improve the communication efficiency of UAV networks, which is proved in the simulation part.
- Second, the ABC routing delineates the consensus scope for the blockchain, namely the candidate relay drones. This method can avoid all drones from participating in the consensus, thus, reducing the communication overhead.

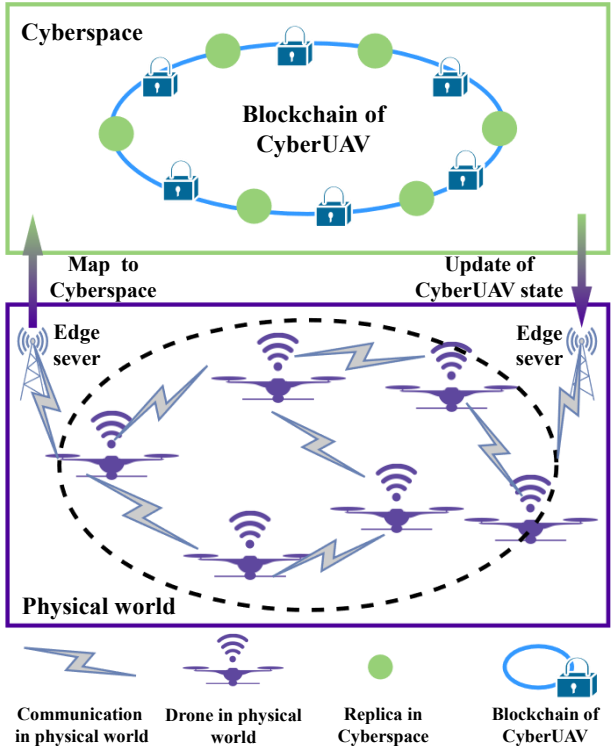

**Fig. 5.** Construction of CyberUAV.

### B. Construction of CyberUAV

The framework of CyberUAV is shown in Fig. 5. During the construction of CyberUAV, three parts, including the *physical world*, *Cyberspace*, and *link between the physical world and Cyberspace*, are involved.

*Physical World:* The physical world includes all physical entities, such as drones and edge servers (ESs). This paper only considers the security of communication between UAVs, and does not consider the communication security between ESs and UAVs or between different ESs, because this part has been already studied extensively [38, 39], and it is beyond the scope of our paper. According to the location distribution of drones in the UAV network, drones in a region are managed by a group of ESs, which are responsible for mapping the UAV network in the physical world to Cyberspace and updating the blockchain information of CyberUAV to the physical world.

*Cyberspace:* Corresponding to the physical world, Cyberspace is a virtual world. It contains digital replicas of drones, and this virtual network built from those replicas is called CyberUAV. Meanwhile, we explore building a blockchain system in Cyberspace, so we further develop a private blockchain structure for CyberUAV because of the high security and access control mechanism for the private chain.

*Link between Physical World and Cyberspace:* In the ESCM framework, the ES is designed as a bridge between the physical world and Cyberspace. The drone in the physical world will first choose an ES within its communi-cation range and build its replica in Cyberspace. In this way, the replica can be located on the ESs to find the corresponding drone. Meanwhile, an ES can operate multiple replicas of drones. In contrast to traditional blockchains relying on P2PC, the communication method in the blockchain of CyberUAV between replicas is V2VC, which is not limited by physical channels. In other words, the communication between replicas can be completed by inter-process communication (IPC), which greatly reduces the latency.

In addition, there are also two steps involved in the construction of the CyberUAV: *mapping* and *updating*.

*Map to Cyberspace:* The ES maps the identity information, private attributes and roles of all drones in the administrative area to corresponding replicas in Cyberspace in the form of DT. The information about each drone can be expressed as

$$replica_i = \{sig(ID), Atr, Role\}, \quad (16)$$

where $sig(ID)$ represents the signature of the drone's identity. $Atr$ is the drone's private attributes, such as coding ability, and communication ability. And $Role$ represents the role played by this drone, such as the sender drone, the receiver drone, or the drone as the candidate relay node.

Furthermore, each replica encapsulates its update log into a network transaction, and broadcasts the transaction for other replicas to reach consensus with a format generally described as

$$Tx_i = \{replica_i, sig(req), Adr, Ts\}, \quad (17)$$

where $sig(req)$ represents the signature of the transaction request. $Adr$ is two addresses which are the addresses of the sender drone and the receiver drone. And $Ts$ is the timestamp of this transaction.

*Update of CyberUAV state:* After a consensus is reached on the blockchain of CyberUAV in Cyberspace, one of the candidate drones is selected as the final relay node to route the message sent by the sender drone and forward them to the receiver drone. At this time, this consensus result should be notified by the corresponding ES to the UAV network in the physical world to complete the above routing step. In addition, this ES should also update the CyberUAV block generation on the corresponding UAV in the physical world.

After the introduction of the DT, ESCM gains two advantages:
- First, the DT can achieve a stable connection of blockchain, and increase the consensus rate;
- Second, DT also avoids P2PC to reduce communication delay.

### C. Network Coding

Through the mapping of the sender drone, the receiver drone and candidate relay drones from the UAV network, CyberUAV forms a special static network named combined network (CN) for CyberUAV, because CN has advantages in coding and cost [40]. Before introducing the blockchain consensus PoNC based on network coding, we need to clarify our coding scheme for the CN.

The CN is a special multicast network whose structure is similar to a multi-fork tree. The network is divided into three layers, which are composed of the sender node (according to the sender drone), relay nodes (according to the relay drones), and receiver nodes (according to the receiver drones), successively. In our ESCM, the number of sender nodes and relay nodes is 1 and $n$, respectively, and the number of receiver nodes is $m$ (it depends on how many receiver nodes the sender node needs to send messages to, because in practice the sender node may send messages to different receiver nodes). Meanwhile, each receiver node connects to $k$ relay nodes (it can be determined by routing protocol). In general, the value of $k$ is less than the value of $n$, because $n$ relay nodes may contain relay nodes corresponding to other receiver nodes. The expression of the topological structure of CN is $C_n^k$. Fig. 6 shows the CN topological structure of $C_3^2$, $C_4^2$ and $C_4^3$ respectively.

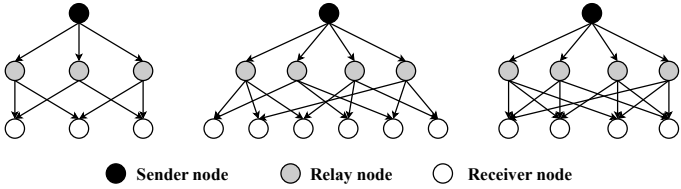

**Fig. 6.** Network topology of CN (From left to right are $C_3^2$, $C_4^2$ and $C_4^3$).

According to the theory of CN, the encoding vector of $k$ messages from the $k$ relay nodes is linearly independent and can be decoded directly, thus reducing the complexity of decoding [41]. In CN, the sender node assigns coding vectors $(1, a_1), (1, a_2), \ldots, (1, a_k)$ to $k$ relay channels corresponding to relay nodes to realize encoding and forwarding the original messages. Due to a receiver node may receive encoded messages forwarded by different relay nodes, it needs to determine which relay node forwards the message, in order to decode it. When a receiver node receives messages forwarded by different relay nodes, it only needs to judge the linear correlation of multichannel coding vectors to find the corresponding relay node for decoding [41].

When $k=2$ the CN coding is carried out for $C_n^2$, and the corresponding coding matrix vector $\begin{bmatrix} 1 & \alpha \\ 1 & \beta \end{bmatrix}$ is established for any two vectors $(1,\alpha)$ and $(1,\beta)$, then the determinant of the matrix is

$$\begin{vmatrix} 1 & \alpha \\ 1 & \beta \end{vmatrix} = \beta - \alpha. \qquad (18)$$

If the determinant of (18) is not 0 and $\alpha \neq \beta$, then the two vectors $(1,\alpha)$ and $(1,\beta)$ are linearly independent.

When $k=3$ the CN coding is carried out for $C_n^3$, and the corresponding coding matrix vector $\begin{bmatrix} 1 & \alpha & \alpha^2 \\ 1 & \beta & \beta^2 \\ 1 & \gamma & \gamma^2 \end{bmatrix}$ is established for any three vectors $(1,\alpha,\alpha^2)$, $(1,\beta,\beta^2)$ and $(1,\gamma,\gamma^2)$, then the determinant of the matrix is

$$\begin{vmatrix} 1 & \alpha & \alpha^2 \\ 1 & \beta & \beta^2 \\ 1 & \gamma & \gamma^2 \end{vmatrix} = \beta\gamma^2 + \alpha\beta^2 + \alpha^2\gamma - \alpha^2\beta - \alpha\gamma^2 - \beta^2\gamma \qquad (19)$$

$$= (\beta-\alpha)(\gamma-\beta)(\gamma-\alpha).$$

If the determinant of (19) is not 0 and $\alpha \neq \beta \neq \gamma$, then the three vectors $(1,\alpha,\alpha^2)$, $(1,\beta,\beta^2)$ and $(1,\gamma,\gamma^2)$ are linearly independent.

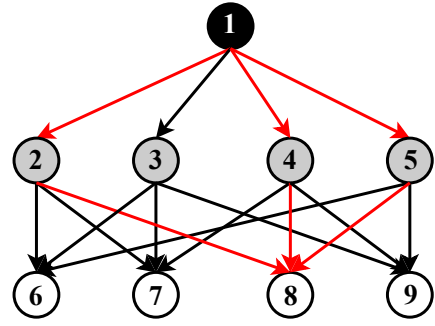

**Fig. 7.** Forwarding paths of the message in $C_4^3$ network.

Then, when carrying out CN coding for $C_n^k$, the receiver node should receive $k$ messages from different relay nodes, and the corresponding $k$ vectors are $(1, a_1, a_1^2, a_1^3, \ldots, a_1^{k-1})$, $(1, a_2, a_2^2, a_2^3, \ldots, a_2^{k-1})$, $\ldots$, $(1, a_k, a_k^2, a_k^3, \ldots, a_k^{k-1})$ to establish the coding vector matrix $\begin{bmatrix} 1 & \cdots & a_1^{k-1} \\ \vdots & \ddots & \vdots \\ 1 & \cdots & a_k^{k-1} \end{bmatrix}$. And the determinant of this matrix is

$$\begin{vmatrix} 1 & \cdots & a_1^{k-1} \\ \vdots & \ddots & \vdots \\ 1 & \cdots & a_k^{k-1} \end{vmatrix} = \prod_{1 \leq i < j \leq k} (a_i - a_j). \qquad (20)$$

If the value of the determinant (20) is not 0 and $a_1 \neq a_2 \neq \ldots \neq a_n$, the $k$ messages received by the receiver node are linearly independent.

### D. PoNC Consensus

As mentioned above, the ESCM selects multiple candidate relay nodes for a sender node through the ABC routing protocol. As shown in Fig. 7, when the *sender node* 1 sends *message A* to the *receiver node* 8, it can pass through three paths of three trunk nodes, namely (21). Therefore, we need to use an appropriate consensus algorithm to determine the best relay node to encode and forward *message A*.

$$< 1 \rightarrow 2 \rightarrow 8 >,$$
$$< 1 \rightarrow 4 \rightarrow 8 >, \qquad (21)$$
$$< 1 \rightarrow 5 \rightarrow 8 >.$$

Since the purpose of selecting nodes is to find an appropriate relay node to encode *message A*, we consider the encoding ability of a candidate relay node as the evaluation factor. In addition, coding and mining are similar in that both require a certain amount of computing resources, so we chose to design our PoNC consensus according to PoW consensus.

Although there are other consensus algorithms designed according to PoW, such as PoSo [42], there is no consensus algorithm based on network coding at present, and network coding can significantly improve network throughput, thus improving the communication efficiency of UAV networks. For PoSo, it successively selects each node to provide problem solutions, which are verified by the other nodes to determine the best solution and the master node. Finally, the solution approved by more than 50% of nodes is determined as the execution basis. This verification process takes too long and is not suitable for high-speed and dynamic UAV networks.

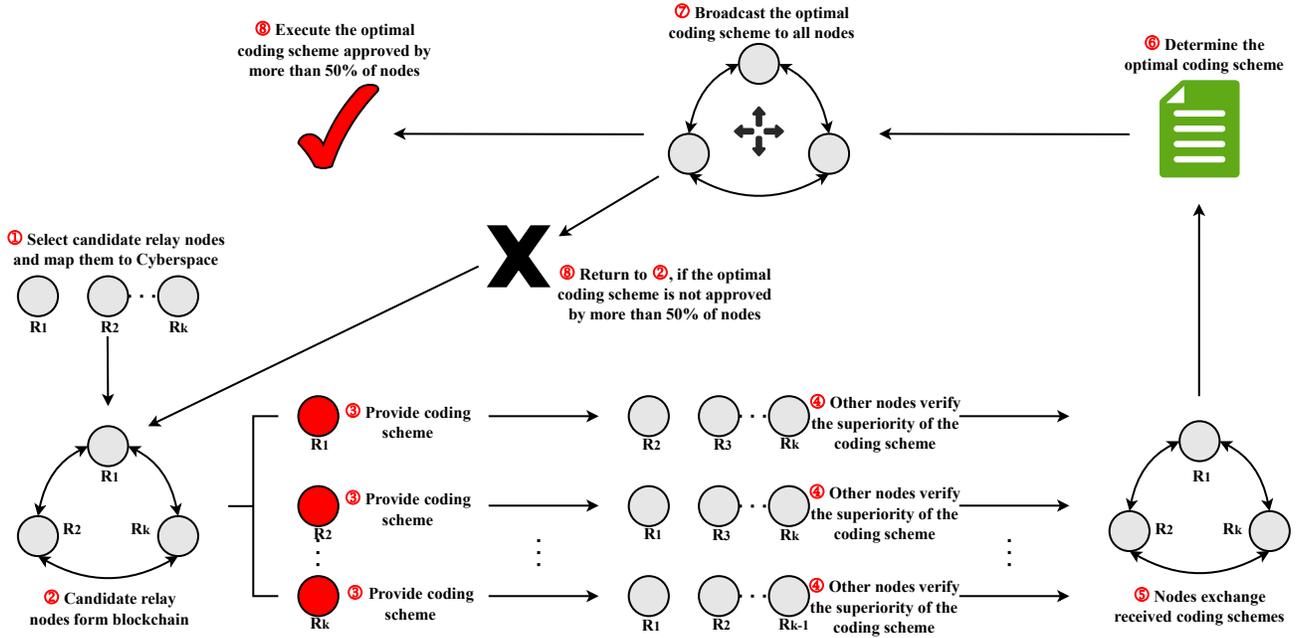

**Fig. 8.** Flow chart of PoNC.

Our PoNC can replace meaningless mining behavior in PoW with meaningful coding behavior. Similar to the concept of "miners" in PoW, PoNC imagines the relay nodes that play: the role of encoding and forwarding messages as "coders". Meanwhile, the coding capability of each candidate relay node is verified in parallel, which saves the verification time and improves the efficiency of the entire ESCM. The specific consensus process is shown in Fig. 8, the steps are as follows.

Step (1). Candidate relay nodes are selected by the ABC routing protocol, and mapped to Cyberspace by ESs.

Step (2). These candidate relay nodes form a blockchain system CyberUAV in Cyberspace.

Step (3). Each node provides its coding scheme for forwarded messages. It should be noted that the actions of each node occur simultaneously and in parallel.

Step (4). The other nodes verify the coding scheme given by a node. For example, if the coding scheme comes from candidate relay node $R_1$, the remaining candidate relay nodes $R_2, R_3, \ldots, R_k$ will verify its superiority.

Step (5). When the verification is complete, all nodes exchange their views on every coding scheme with each other.

Step (6). According to the view of all nodes, they determine the optimal coding scheme.

Step (7). When the optimal coding scheme is determined, the provider of the scheme broadcasts the optimal coding scheme to all nodes in the blockchain.

Step (8). When more than 50% of nodes approve the coding scheme after broadcasting, the provider of the scheme is determined as the final relay node and performs the coding scheme for the forwarded message. If the scheme is not supported by more than 50% of nodes, consensus will return to Step (2) and select a new coding scheme.

The 50% threshold of PoNC is from PoW [24]. Since network coding is an optimization problem [43], the specific process of the verification method in Step (4) is similar to that in [42]. In addition, if a malicious node exists, the exchange of the verification results in Step (5) will be affected. The circumvention of this problem is also partially included in [42]. Furthermore, we also theoretically present the security analysis of two adversary models in Section VI-B, and the simulation in Section VII-D.

PoNC can bring the following benefits to the ESCM:
- First, PoNC selects relay drones with the best coding capability. In the consensus process, not only relay nodes are selected, but also the coding scheme is determined;
- Second, the coding is a meaningful behavior in the network, which avoids a large amount of computational power waste caused by PoW consensus. Although coding schemes from some candidate relay drones are not adopted, they may be used as a backup when the selected relay drone fails. Additionally, these drones may also serve as the best relay nodes for other receiving drones, thus, this coding scheme can also play a role, indicating that computing resources are not wasted;
- Third, coding can further improve the throughput of the UAV network.

## VI. PERFORMANCE ANALYSIS

In this section, we conduct a theoretical analysis of PoNC consensus, including consensus overhead and security.

### A. Consensus Overhead

Firstly, we model the overhead in PoNC consensus. Due to the fact that drones with high-speed states are particularly sensitive to delay, the overhead here refers to the communication cost in the consensus process.

For PoNC, from Step (3) to Step (4), each node needs to inform other nodes of its coding scheme, so the communication cost is ($k$-1). Since there are $k$ nodes providing the coding scheme, the communication cost of this process is

$k(k-1)$. In addition, in Step (5), each node needs to exchange the received coding scheme, which requires all nodes to communicate in pairs, so the communication cost is $k^2$. Finally, after the optimal coding scheme is determined, it needs to be broadcast to the whole network by the scheme provider in Step (7), and the communication cost is $(k-1)$. As a result, the communication overhead of PoNC is

$$C_{PoNC} = k(k-1) + k^2 + (k-1) = 2k^2 - 1. \quad (22)$$

Secondly, we evaluate the communication overhead of the consensus algorithm PoSo [42] also based on PoW. Since the verification process of PoSo is not parallelized, it is necessary to verify the superiority of the scheme provided by each node in turn. In the worst case, we need to verify all the schemes of $k$ nodes in turn, and in the best case, the scheme of the first node is the optimal solution. We select the mean value, that is, verifying half of the nodes $k/2$ to find the optimal solution, and its communication overhead is

$$C_{PoSo} = \frac{k}{2}\left[(k-1) + (k-1)^2 + (k-1)\right] = \frac{k^3}{2} - \frac{k}{2}. \quad (23)$$

### B. Consensus Security

Consistent with the adversary model we described in Section III, we consider two types of consensus security issues.

*1) External Attack*

Assume that the probability of MNs generating a block is $p_m$, and the probability of HNs producing a block is $p_h$. According to the operation mechanism of blockchain, when the probability of MNs generating blocks is bigger than the probability of HNs producing blocks, DSA is a successful attack, and the probability $P_{S\_E}$ is

$$P_{S\_E} = \begin{cases} 1 & , p_h \leq p_m \\ \left(\dfrac{p_m}{p_h}\right)^z & , p_h > p_m \end{cases}, \quad (24)$$

where $z$ represents the number of blocks with more HNs than MNs. With the continuously increasing number of blocks, the probability of a successful attack will decrease exponentially, which is Poisson distribution, and the expected value can be expressed as

$$\alpha = z \frac{p_m}{p_h}. \quad (25)$$

Then, the probability of a successful attack can be expressed as the probability density of (24) times (25)

$$P_{S\_E} = \sum_{i=0}^{+\infty} \frac{\alpha^i e^{-\alpha}}{i!} \begin{cases} 1 & , i > z \\ \left(\dfrac{p_m}{p_h}\right) & , i \leq z \end{cases}. \quad (26)$$

The above equation can also be simplified as

$$P_{S\_E} = 1 - \sum_{i=0}^{z} \frac{\alpha^i e^{-\alpha}}{i!}\left(1 - \left(\frac{p_m}{p_h}\right)^{(z-i)}\right). \quad (27)$$

*2) Internal Attack*

After the drone is mapped to the replica node, the node's identity will not change in CyberUAV, it means that the honest drone will not become an MN. Therefore, the presence of MNs in CyberUAV may be MDs in the candidate relay drones selected by the ABC routing. However, it is also possible that we map MDs into CyberUAV.

Thus, we assume that there are $N$ drones in the UAV network, and $X$ of them are MDs. The proportion of MDs to all drones is $R$. When the ABC routing selects $k$ candidate relay drones, there are $x$ MDs among them. This selection process can be regarded as a sample survey of products, and the probability can be expressed as (28), which is a hypergeometric distribution.

$$P(x) = \frac{C_X^x C_{N-X}^{k-x}}{C_N^k}. \quad (28)$$

According to PoW, an attacker needs to control 50% or more of the computing resources to change consensus results by internal attack [24]. Similarly, in order to change the consensus result of PoNC by an internal attack, condition (29) needs to be satisfied.

$$P_{S\_I} = \sum_{i=1}^{x}\left(P(x), \text{ when } \sum_{i=0}^{x} CR_i \geq 50\% \sum_{j=0}^{k} CR_j\right). \quad (29)$$

where $CR$ represents the computing resources of each node.

## VII. PERFORMANCE EVALUATION

In this section, we simulate the performances from many areas, including the ABC routing protocol, CN coding scheme, PoNC consensus, and MobileChain.

### A. Simulation Setup

We build the simulation environment in OMNet++ (an open-source network simulation software), as shown in Fig. 9, and the simulation parameters are shown in Table. I.

### B. ABC Routing Protocol

The routing protocol is the key technology for selecting candidate relay drones, and its time determines the efficiency of the ESCM building blockchain. Therefore, this part simulates the ABC routing to test the communication delay and message arrival rate of candidate drones when forwarding messages under different situations (such as different numbers and flight speeds of drones), so as to verify the reliability of the candidate relay drones (we set $k=5$ here) found by the ABC routing protocol. We use the average value of the performance of these drones to compare with other works, namely ABNT [7], and LB-OPAR [45].

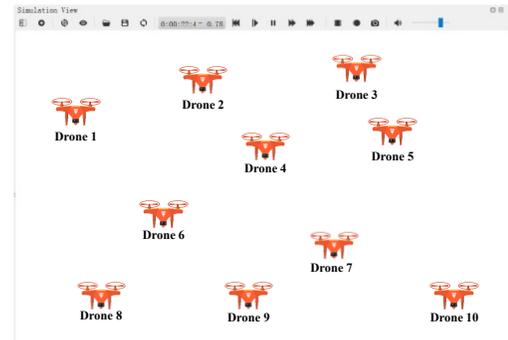

**Fig. 9.** Simulation environment (Take ten drones as an example).

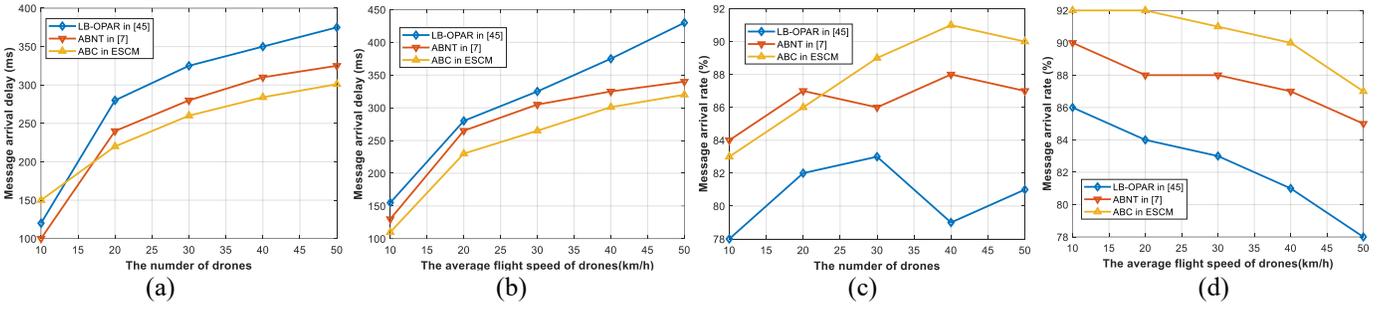

**Fig. 10.** (a) Message arrival delay with the different number of drones ($v$=40 km/h). (b) Message arrival delay with the different average flight speeds of drones ($N$=50). (c) Message arrival rate with the different number of drones ($v$=40 km/h). (d) Message arrival rate with the different average flight speeds of drones ($N$=50).

TABLE I
Parameters of Simulation

| Parameter | Value |
|---|---|
| Covered area | 1km*1km*1km |
| MAC layer protocol | IEEE 802.15.4 [44] |
| Communication distance of a drone | 20 m |
| Frequency of generating BM | 5 Hz |
| Life period of BM | 15 hops |
| BM size | 16 bits |
| Message size | 128 bits |
| Interval between each message | 1 s |
| *limit* | 20 |
| *MaxGen* | 500 |

Fig. 10 (a) shows the relationship between the message arrival delay and the number of drones when the average flight speed is 40km/h. In general, delay increases with the increase of the number of drones. In addition, when the number is less than 20, the ABC routing is inferior to ABNT in [7], and LB-OPAR in [45]. The main reason is that we select 5 candidate relay nodes and measure their average delay. When the scale of drones is small, the selection range of candidate relay nodes for the ABC routing is small, resulting in poor average performance. However, when the number of drones is greater than 20, the ABC routing has obvious advantages in terms of delay.

Fig. 10 (b) shows the relationship between the message arrival delay and the average flight speed of drones when the number is 50. The result shows that as the speed of the drone increased, so does the delay. It also shows that the ABC routing in ESCM is superior to the methods of [7] and [45].

Fig. 10 (c) shows the relationship between message arrival rate (the number of messages received to the number of messages sent) and the number of drones when the average flight speed is 40km/h. This index fluctuates with the increase of the number of drones. In addition, when the number is less than 20, the ABC routing is slightly worse than the method in [7]. When the number of drones is more than 20, the ABC routing is the best among the three. We believe that the main reason for this phenomenon is consistent with Fig. 10 (a).

Fig. 10 (d) shows the relationship between the message arrival rate and the average flight speed when the number of drones is 50. Generally, the message arrival rate decreases with the increase of average flight speed. However, the ABC routing in the ESCM is superior to the other two comparison methods.

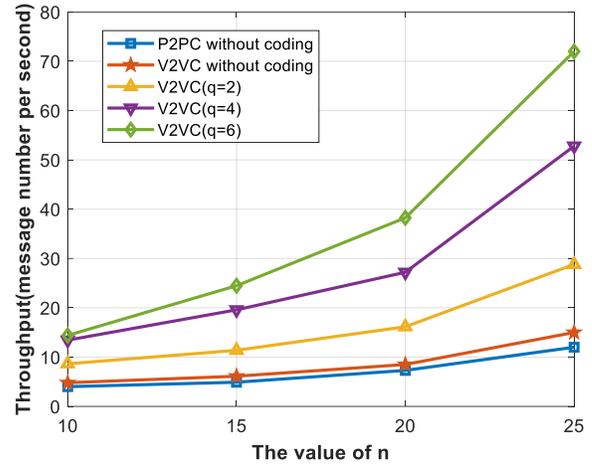

**Fig. 11.** Throughput.

*C. CN Coding Scheme*

This part simulates the impact of the CN encoding scheme on the throughput. The simulation parameters and platform are the same as part *A*. This simulation envisages such a scenario, when a receiving node needs to receive $q$ messages, the best scheme is for the PoNC consensus algorithm to select $q$ relay nodes from $k$ candidate relay nodes to forward messages in parallel. It can avoid the delay of queue forwarding messages successively. Then, we explore the throughput changing law by varying the values of $n$ and $q$. It is important to note that $q<k<n$. Additionally, we also compare the throughput difference between P2PC and V2VC in the simulation.

According to Fig.11, we can find that the V2VC has an advantage over P2PC in terms of throughput without coding, because the V2VC removes the communication delay at the physical layer and only preserves the network layer communication, thus improving the throughput. In addition, when we code for V2VC, the throughput can be positively affected by the increase of $q$ and $n$ values. The reason why $q$ has a positive impact on throughput is that the larger $q$ is, the more ability of network coding can be used. This rule implies that we can consider accumulating multiple delay-insensitive messages to be forwarded parallelly in practice. Moreover, the value of $n$ is affected by both $k$ and $m$ values, indicating that we can consider multiple receiver nodes at the same time, and increase the number of candidate nodes $k$ selected in ABC routing, to improve the network throughput. The above simulation results and laws are consistent with the theory in [40]. In general, the V2VC and the network coding can simultaneously improve the throughput of UAV networks.

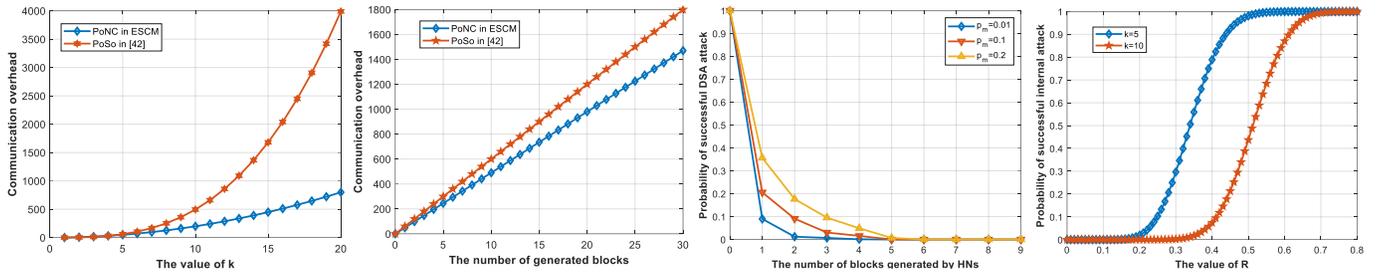

**Fig. 12.** (a) Communication overhead with the different values of $k$. (b) Communication overhead with the different number of blocks ($k$=5). (c) Probability of successful DSA. (d) Probability of successful internal attack.

### D. PoNC Consensus

In this part, we simulate PoNC from two aspects, which are consensus overhead and security, so as to prove its superiority.

*1) Consensus Overhead*

Fig.12 (a) compares the communication overhead of PoNC and PoSo [42]. It shows the relationship between the number of consensus nodes and the communication overhead of PoNC and PoSo. When the number of consensus nodes increases, PoNC has a significant advantage over PoSo in terms of communication overhead. This result shows the potential scalability of PoNC.

Furthermore, we set the value of $k$ to 5, that is, five candidate relay nodes participate in consensus, to simulate the relationship between the communication overhead of the two consensuses and the number of generated blocks. Fig.12 (b) shows this result, indicating that the communication overhead of the two consensus algorithms increases linearly. However, the communication overhead of PoNC is always lower than that of PoSo.

*2) Consensus Security*

In consensus security analysis, we divide the security into external and internal security for the proposed adversary model. External security is facing external attackers while internal security is solving internal malicious nodes.

*External Attack:* In the DSA model, MNs do not affect HNs generating blocks. Therefore, HNs generate blocks according to the normal consensus process, and the probability $P_h$ defaults to 1. However, the probability of MNs generating blocks is closely related to the number of MNs and computing resources, which can be used as the dependent variable of (27) to analyze the probability of a successful attack. Thus, do numerical simulations of this metric with $p_m$=0.01, $p_m$=0.1, and $p_m$=0.02.

Fig.12 (c) shows the relationship between the probability of a successful attack and the number of blocks generated by HNs under different probabilities of MNs generated blocks. It can be found that the probability of a successful attack decreases significantly with the increase of the number of blocks. When the number is greater than 5, it is almost impossible for MNs to successfully attack.

In addition, the consensus process is conducted in Cyber-UAV, a private space we established in Cyberspace. Due to the strict entry mechanism of the private space, the probability of MNs existence is very small, and the probability of block generation is far less than that of HNs, so PoNC can effectively resist DSA. Meanwhile, each communication from the UAV network creates a block via consensus. Obviously, there are a lot of communications in the UAV network, and the number of blocks will also be large. Then, according to the simulation results, it is impossible for MNs to complete the attack when the number of blocks reaches a certain level, which proves the security of PoNC.

*Internal Attack:* Then we set the value of $CR$ as the Gaussian distribution between the range [100 kbps, 300 kbps], and $N$=50. Furthermore, we simulate this metric to obtain Fig. 12 (d).

Fig. 12 (d) shows the relationship between the probability of a successful internal attack and the value of $R$. The bigger the proportion of MNs, the easier an internal attack is to achieve. However, as the number of candidate relay nodes increases, this kind of internal attack can be resisted and the probability of a successful attack can be reduced. Therefore, PoNC is secure enough against internal attack.

### E. Ablation Study

Finally, we conduct an ablative study to verify that each module is effective for ESCM. We use message arrival delay and message arrival rate as evaluation metrics. Furthermore, we take the performance of the ABC routing protocol as the baseline. Similar to Part A of this section, we fix the UAV flight speed $v$=40 km/h, and the UAV number $N$=50 to evaluate these two performances.

The simulation results are shown in Fig. 13. Compared with the baseline, our proposed ESCM has extremely high advantages in both efficiency and security. Moreover, the reduction of any module will lead to the degradation of ESCM performance, indicating that network coding, blockchain consensus and DT are all necessary and important for our ESCM.

For the **massage arrival delay**, the most serious impact on this metric is network coding, followed by PoNC consensus and DT. This indicates that network coding performance, as the validation index of blockchain consensus, plays an important role in improving the communication efficiency. Meanwhile, blockchain consensus can further determine the node with the best coding ability to further improve communication efficiency. Additionally, DT can use V2VC to avoid the delay caused by P2PC.

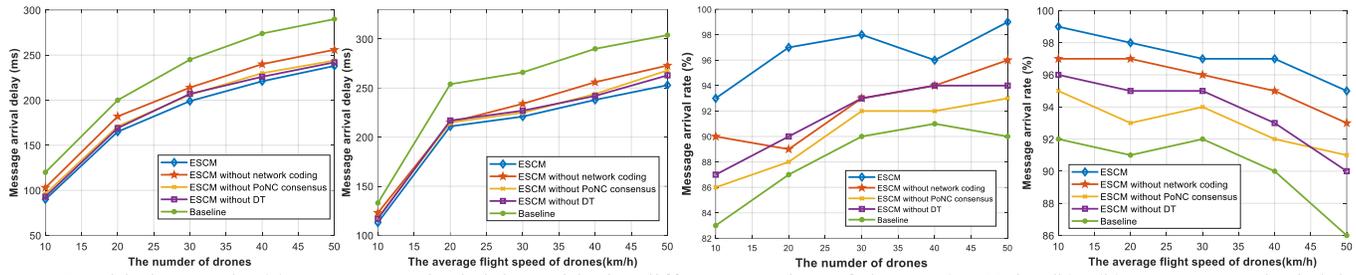

**Fig. 13.** Ablation study (a) message arrival delay with the different number of drones (*v*=40 km/h). (b) message arrival delay with the different average flight speeds of drones (*N*=50). (c) message arrival rate with the different number of drones (*v*=40 km/h). (d) message arrival rate with the different average flight speeds of drones (*N*=50).

For **massage arrival rate**, PoNC consensus has the most serious impact on this metric, followed by DT and network coding. This shows that blockchain has a significant role in ensuring communication security, and the network can achieve consistency through consensus. In addition, DT can construct static Cyberspace for wireless mobile networks and provide a stable communication environment for PoNC consensus, thus improving message arrival rate.

## VIII. Conclusion

In this paper, through a routing protocol, DT technology, network coding and blockchain consensus, ESCM provides a complete solution for efficient and secure communication in UAV networks. First, we demonstrate that MobileChain has a low consensus success rate due to its low communication success rate, illustrating the superiority and necessity of blockchain consensus using DT-enabled in wireless mobile networks. Then, we introduce the roles ABC routing, DT, network coding and PoNC consensus play in ESCM and their design details in turn.

The analysis and simulation results show the superiority of the above modules. In particular, ablation studies confirm that these modules are necessary and useful for ESCM.

In general, ESCM has advantages in different performance aspects and can be well adapted to high-speed mobile environments such as UAV networks. It should be noted that ABC routing, DT, CN coding and PoNC consensus in ESCM can be used separately and flexibly, rather than necessarily bundled together. In the future, we will explore the performance of ESCM in other network scenarios.